# Fiber Bragg Grating Based Thermometry

Zeeshan Ahmed[1*], James Filla[2], William Guthrie[3] and John Quintavalle[2]

[1]*Thermodynamic Metrology Group, Sensor Science Division, Physics Laboratory, National Institute of Standards and Technology, Gaithersburg, MD 20899*

[2] *Innovations and Solutions Division, National Institute of Standards and Technology, Gaithersburg, MD 20899*

[3] *Statistical Engineering Division, Sensor Science Division, Physics Laboratory, National Institute of Standards and Technology, Gaithersburg, MD 20899*

**Abstract:**

In recent years there has been considerable interest in developing photonic temperature sensors such as the Fiber Bragg gratings (FBG) as an alternative to resistance thermometry. In this study we examine the thermal response of FBGs over the temperature range of 233 K to 393 K. We demonstrate, in hermetically sealed dry Argon environment, FBG devices show a quadratic dependence on temperature with expanded uncertainties ($k=2$) of ≈500 mK. Our measurements indicate that the combined measurement uncertainty is dominated by uncertainty in determining peak center fitting and thermal hysteresis of polyimide coated fibers.

---

[*]Contact author: zeeshan.ahmed@nist.gov

**Introduction:**

Temperature is one of the most measured quantities in the world. Despite the ubiquity of thermometers, the underlying technology has been slow to advance over the last century.[1] For the past century, industrial temperature measurements have relied on resistance measurement of a thin metal wire or filament whose resistance varies with temperature. [1, 2] Though resistance thermometers can routinely measure temperatures with uncertainties of ≤10 mK (US industry requirement), they are sensitive to mechanical shock induced strain which causes the sensor resistance to drift over time.[2] Consequently, resistance thermometers require frequent off-line, expensive, and time consuming calibrations resulting in high ownership cost. Additionally, these devices suffer from electromagnetic interference, limiting their utility in extreme environments. These fundamental limitations of resistance devices have produced considerable interest in the development of photonic temperature sensors as an alternative to resistance thermometers as they have the potential to leverage advances in frequency metrology to provide greater temperature sensitivity while being robust against mechanical shock and electromagnetic interference.[3-10]

The vast and varied application landscape for temperature measurement has spawned a host of photonic temperature sensing solutions. The proposed sensor technologies range from temperature sensitive dyes [11], polymers [9, 12, 13] to silicon photonics such as ring resonators[4, 5, 7]. Fiber Bragg grating (FBG) based temperature sensors have already been commercially introduced as photonic alternative to resistance thermometry.[14, 15] FBG are a narrow band filter commonly used in the telecommunications industry for routing information. FBGs are commonly fabricated using photo-sensitive optical fibers that are exposed to spatially varying light source such as a deep UV laser in a Michelson interferometer. [14, 15] The periodic varying light induces photo-chemical reactions which modifies the local structure of the fiber to create a periodic variation in the local refractive index that acts like a Bragg grating. Wavelength of light resonant with the Bragg period is reflected back, while non-resonant wavelengths pass through the grating. Change in surrounding temperature impacts the effective grating period by either linear thermal expansion of the material and/or a change in the fiber's refractive index due to temperature (thermo-optic effect).[10, 14, 15] Existing literature indicates the FBG show a temperature dependent shift of ≈10 pm/K around 293 K.[10] There is some disagreement over the temperature dependent behavior and measurement uncertainties with different reports suggesting the thermal response may be linear[16, 17] or quadratic[18].

In this study we have examined the temperature dependent response of polyimide coated FBGs over the temperature range of 233 K to 393 K against a calibrated Platinum resistance thermometer (PRT). Our results indicate that the FBG show a quadratic dependence on temperature with combined expanded measurement uncertainties of ≈500 mK (*k=2*).

**Experimental:**

*Fiber Bragg Gratings:* In this study we have utilized commercially available polyimide coated silica fiber based FBG with Bragg resonance set at 1540 nm, 1550 nm (3 fibers), and 1560 nm. Fibers where stored in a humidity controlled environment (20 % RH) prior to use. Each fiber was cleaved such as to leave 5 mm of excess fiber on one side of the sensor, with the other side, 0.5 m long terminated in a fiber optic coupler. The sensor was then guided through a 23G needle (inserted through a cork) into a 240 mm long (6.5 mm diameter) glass tube. The bottom 50 mm of glass tube was filled finely ground, dry MgO power, completely immersing the sensor, to ensure excellent thermal contact between the temperature bath and FBG. Packing the sensor in MgO powder improves the temperature repeatability by a factor of four (data not shown). A small amount of desiccant was added on top of MgO powder to keep the assembly dry. The glass tube was then backfilled with Ar gas, following which the glass tube was sealed by inserting the cork and sealing the cork/needle/glass interface with quick-dry epoxy.

*FBG interrogation system:* We have custom built a laser based FBG interrogation system for interrogating the FBG. Briefly, the assembled FBG is placed in a cylindrical Al block (25 mm diameter, 170 mm length). The cylinder has two 150 mm long blind holes (2.5 mm and 6.5 mm diameter) for accommodating a calibrated Platinum resistance thermometer (PRT) and the assembled FBG sensor, respectively. The metal block is placed inside the dry temperature bath (Fluke[†]). The dry well temperature is controlled by an automated LabVIEW program which cycles the temperature between 233 K to 393 K at 5 K intervals. Once the set temperature is achieved, the program allows 5 mins of equilibration time following which the laser (New Focus TLB-6700 series) scan is initiated. A small amount of laser power was immediately picked up from the laser output for wavelength monitoring (HighFinesse WS/7) while the rest, after passing through the photonic device via an optical circulator (ThorLabs), was detected by a large sensing-area power meter (Newport, model 1936-R). Five consecutive scans were recorded at each temperature and each sensor was thermally cycled four to five times in each run unless noted otherwise. The recorded data was fitted using a non-parametric fitting routine to extract peak center, peak height and peak width as a function of temperature.

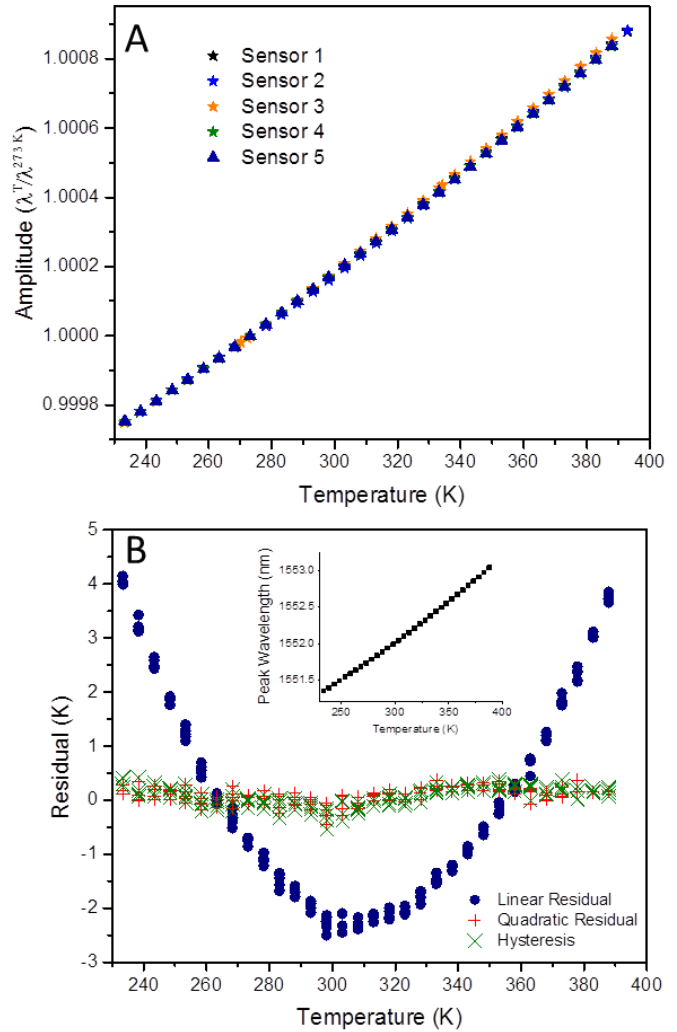

Figure 1: a) Wavelength normalized temperature-dependent response of four different sensors. b) FBG show a quadratic dependence on temperature

---

[†] Disclaimer: Certain equipment or materials are identified in this paper in order to specify the experimental procedure adequately. Such identification is not intended to imply endorsement by the National Institute of Standards and Technology, nor is it intended to imply that the materials or equipment identified are necessarily the best available.

*Temperature dependence of FBG:* The grating equation is given by:

$$\lambda_B = 2n_e L \quad (1)$$

where $\lambda_B$ is the Bragg wavelength, $L$ is the grating period and $n_e$ is effective refractive index. The temperature dependence of the FBG resonance derives from changes in the refractive index ($n_e$) due to the thermo-optic effect and to a lesser extent changes in grating period due to thermal expansion of the fiber. The thermal behavior of FBG can be adequately captured using a Taylor expansion of wavelength as function of temperature which yields the following expression:

$$\lambda^T = \lambda^R + \frac{\partial \lambda}{\partial T}(T-R) + \frac{\partial^2 \lambda}{\partial T^2}(T-R)^2 \quad (2)$$

where, $T$ is the sensor temperature, $R$ is the reference temperature (273.15 K) and $\lambda^T$ and $\lambda^R$ are the Bragg wavelength at temperatures $T$ and $R$, respectively. Using the wavelength-period relationship (eq 1), eq 2 can be re-written as:

$$\lambda(T) = a_1 T^2 + a_2 T + a_3 \quad (3)$$

$$a_1 = 2\left\{\left(\frac{\partial^2 n}{\partial T^2}\right)L + (\partial n/\partial T)\left(\frac{\partial L}{\partial T}\right) + \left(n * \frac{\partial^2 L}{\partial T^2}\right)\right\},$$

$$a_2 = \left(2L\frac{\partial n}{\partial T}\right) + \left(2n\frac{\partial L}{\partial T}\right) - \left(4LR\frac{\partial^2 n}{\partial T^2}\right) - \left(4R\frac{\partial n}{\partial T}\frac{\partial L}{\partial T}\right) - \left(4nR\frac{\partial^2 L}{\partial T^2}\right)$$

$$a_3 = 2\left\{\left(LR^2 * \frac{\partial^2 n}{\partial T^2}\right) + \left(\frac{\partial n}{\partial T}\frac{\partial L}{\partial T} * R^2\right) + \left(nR^2 * \frac{\partial^2 L}{\partial T^2}\right) - \left(2LR * \frac{\partial n}{\partial T}\right) - \left(nR * \frac{\partial L}{\partial T}\right) + \lambda^R\right\}$$

Examination of eq 3 shows that significant departure from linearity in the wavelength-temperature relationship would indicate that the thermo-optic coefficient is varying with temperature. Studies of thermal dependence of thermo-optic coefficient in fused silica[19] indicate a monotonic dependence on temperature.

**Results and Discussion:**

The thermal response of FBG was investigated over the range of 233 K to 393 K. As shown in Fig 1 the wavelength-normalized temperature response of the FBG shows a similar profile over a wide range of Bragg wavelengths. Fitting the thermal response to linear function consistently yields parabolic residuals with deviations as large as ≈6 K. In agreement with Flockhart et al[18] we find that fitting to a quadratic function adequately minimizes the residuals to the point where measurement repeatability dominates the residuals. We do not observe any significant correlation between bandwidth or amplitude and temperature (data not shown).

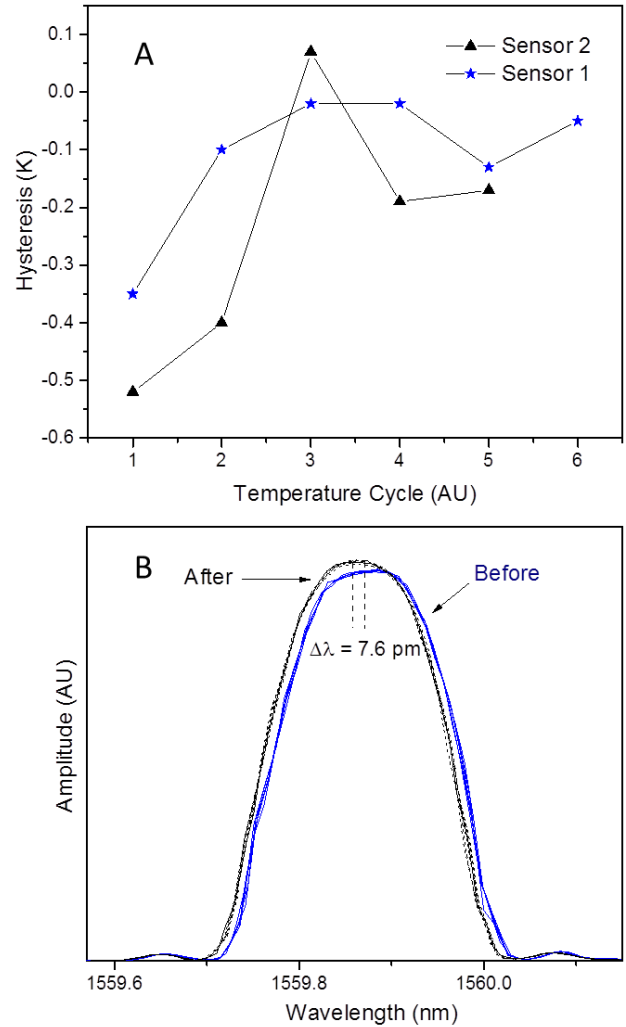

Thermal cycling reveals the FBG sensors undergo a temperature related "aging" process that results in significant hysteresis. As shown in Fig 2a, the thermal hysteresis is largest in the first cycle and steadily decreases over subsequent cycles. We estimate the contribution of hysteresis to measurement uncertainty be taking standard deviation of the mean of hysteresis in each subsequent thermal cycle excluding the first cycle. Annealing or aging of the fibers is only observed when fiber is heated to temperatures above 373 K (data not shown). Examination of the temperature induced drift indicates that annealing at temperatures of 373 K or above results in a clear downshift of the Bragg resonance at 293 K indicating the effective path length of the grating has decreased. The 7.6 pm downshift corresponds to temperature drift of ≈ 0.76 K. This change likely derives from changes in the crystallinity of polyimide.[20]

Figure 2: a) polyimide coated FBG sensors show a thermal induced hysteresis that reduces with increasing runs c) heating sensor to 393 K results in a 6 K drift.

Once the FBG have been properly annealed, the measurement uncertainty in our measurements is dominated by the uncertainty in thermal hysteresis and best fit uncertainty which includes contributions from peak center determination (Table 1). At the laser powers used in the study (170 μW), laser induced self-heating is an insignificant source of uncertainty (data not shown) as is humidity, which if not properly controlled for can be a significant source of uncertainty[21]. As shown in Table 1, for five thermally annealed fibers, our combined expanded measurement uncertainty for humidity controlled, strain-free sensor is ≈ 500 mK.

| Table 1: Uncertainty in FBG temperature measurement (all values in degrees Kelvin) | |
| --- | --- |
| *Temperature (PRT)* | 0.002 |
| *Wavelength* | 0.01 |
| *Fit Residual* | 0.14 |
| *Hysteresis* | 0.2 |
| *Combined Uncertainty (k=1)* | 0.24 |
| *Expanded Uncertainty (k=2)* | 0.49 |

**Summary**: Using careful, traceable measurements of FBG temperature sensors in controlled environment we have demonstrated that FBG sensors can be used to make accurate temperature measurements over the range of 233 K to 393 K. Our results for polyimide coated sensors indicate the fibers undergo temperature induced structural changes resulting in significant hysteresis that reduces over subsequent cycles. Hermetically sealed, annealed sensors, in low humidity and strain-free environment show expanded measurement uncertainties of ≈500 mK.

**Acknowledgement:** The authors would like to acknowledge Arec Jamgochian and Wyatt Millar for help in setting up measurements and Kevin Douglass for helpful discussions.